\documentclass[pre,twocolumn,superscriptaddress]{revtex4}
\usepackage{graphicx}
\usepackage{amsmath,mathrsfs}
\usepackage{braket,enumerate}
\usepackage{amssymb,bm}
\usepackage{bm,hyperref}
\usepackage{CJK,color}

\begin{document}
\begin{CJK*}{UTF8}{gbsn}
\title{Exact Equivalence between Quantum Adiabatic Algorithm and Quantum Circuit Algorithm}
\author{Hongye Yu(余泓烨)}
\affiliation{International Center for Quantum Materials, Peking University, 100871, Beijing, China}
\author{Yuliang Huang(黄宇亮)}
\affiliation{Department of Radiation Oncology, Peking University Cancer Hospital and Institute, 100142, Beijing, China}
\affiliation{International Center for Quantum Materials, Peking University, 100871, Beijing, China}
\author{Biao Wu(吴飙)}
\email{wubiao@pku.edu.cn}
\affiliation{International Center for Quantum Materials, Peking University, 100871, Beijing, China}
\affiliation{Wilczek Quantum Center, School of Physics and Astronomy, Shanghai Jiao Tong University, Shanghai 200240, China}
\date{\today}

\date{\today}
\def\fl{\mathfrak L}
\begin{abstract}
We present a rigorous proof that quantum circuit algorithm can be transformed into quantum adiabatic algorithm with the exact same time complexity. 
This means that from a quantum circuit algorithm of $L$ gates we can construct
a quantum adiabatic algorithm with time complexity of $O(L)$. 
Additionally, our construction shows that one may exponentially speed up some quantum adiabatic algorithms by properly choosing an evolution path. 
\end{abstract}

\maketitle
\end{CJK*}
\def\fl{\mathfrak L}
\def\be{\begin{equation}}
\def\ee{\end{equation}}
\def\ba{\begin{eqnarray}}
\def\ea{\end{eqnarray}}
\def\sdg{Schr\"odinger }
\def\cb{{\cal B}}
\def\cm{{\cal M}}
\def\cf{{\cal F}}
\def\ch{{\cal H}}
\def\cp{{\cal P}}
\section{Introduction}
Quantum algorithms have  two paradigms,  quantum circuit
algorithm~\cite{ChuangBook} and  quantum adiabatic algorithm~\cite{Farhi2000}.
The latter works by adiabatically evolving in the ground state of a system with Hamiltonian
\begin{equation}
H(s)=(1-s)H_B+sH_P\,,
\label{eq1}
\end{equation}
where $s$ increases with time slowly from 0 to 1. The beginning Hamiltonian $H_B$ has a ground state which is easy to construct and the problem Hamiltonian $H_P$ has a ground state that encodes the solutions of the problem. According to the quantum adiabatic theorem~\cite{Farhi2000,zhang2014hierarchical}, the speed of the algorithm is limited by the minimum energy gap between the ground state and the first excited state during the evolution of $H(s)$. When $H(s)$ has an exponentially small minimum gap, the  algorithm is inefficient.

These two kinds of quantum algorithms are shown to be {\it polynomially} equivalent to each other in terms of time complexity~\cite{Dam1,Dam}.
Here we present a rigorous proof  that any quantum circuit algorithm
can be converted into a quantum adiabatic algorithm with the {\it same}
time complexity.  As it has been shown that a quantum adiabatic algorithm can be converted into
a quantum circuit algorithm with the same time complexity~\cite{Dam1}, 
our result means that quantum circuit algorithm and quantum adiabatic algorithm are exactly 
equivalent to each other.

Here is how the rest of our paper is organized. We first describe the main construction of our algorithm in Section II. In Section III, we  show the details of the algorithm and the physical picture behind it. We conclude in Section IV. Thorough analyses of energy gap and errors of the  algorithm are discussed in Appendix.

\section{Construction of Hamiltonians}
\label{sec:main}
Consider a quantum circuit algorithm that has $n$ qubits and $L$ universal quantum gates,
\be
\ket{\alpha_0}\stackrel{U_1}{\longrightarrow}\ket{\alpha_1}\cdots \ket{\alpha_{\ell-1}}
\stackrel{U_\ell}{\longrightarrow}\ket{\alpha_\ell}\cdots \ket{\alpha_{L-1}}
\stackrel{U_L}{\longrightarrow}\ket{\alpha_L}\,,
\ee
where $U_\ell$ represents the $\ell$th quantum gate operation,
$\ket{\alpha_\ell}=U_\ell\ket{\alpha_{\ell-1}}$. Usually, $\ket{\alpha_0}=\ket{00\cdots0}$.
Our aim is to construct a corresponding quantum adiabatic algorithm that has the same time complexity.
For this purpose, we introduce additional $L$ clock qubits and focus on
a special type of clock states
$
\ket{\ell}^c=\ket{1^\ell 0^{L-\ell}}^c\,,
$
which denotes that the first $\ell$ qubits are ones and the rest  are zeros~\cite{Kitaev,Dam}.
Corresponding to the $\ell$th gate operation, we define an operator
\ba
{\cal O}_\ell &=&\frac{\eta}{2}  I\otimes \ket{\ell-1}^c\bra{\ell-1}^c
- \frac{1}{2}U_\ell\otimes \ket{\ell}^c\bra{\ell-1}^c\nonumber\\
&&- \frac{1}{2}U_\ell^\dagger\otimes \ket{\ell-1}^c\bra{\ell}^c+
\frac{1}{2\eta} I\otimes \ket{\ell}^c\bra{\ell}^c\,,
\ea
where $\eta\ge 1$.  This operator with $\eta=1$ was introduced 
in Ref.\cite{Kitaev,Dam}. We construct the beginning and problem Hamiltonians
\begin{eqnarray}
H_B&=&I\otimes\sum_{l=1}^L\ket{l}^c\bra{l}^c\,;\\
H_P&=&\sum_{\ell=1}^L{\cal O}_\ell\,.
\label{ph}
\end{eqnarray}
The ground state of $H_P$ with $\eta>1$ is
\begin{equation}
\ket{\psi^\eta}=\sqrt{\frac{\eta^2-1}{\eta^{2L+2}-1}}\sum_{\ell=0}^L \eta^{\ell} \ket{\gamma_\ell}\,,
\label{gs}
\end{equation}
where $\ket{\gamma_\ell}=\ket{\alpha_\ell}\otimes \ket{\ell}^c$. The ground state energy is 0.
According to the Gershgorin circle theorem, its first excited state has an energy larger 
than $\frac{1}{2}(1/\eta+\eta)-1$, which is finite and independent of the system size.
If our algorithm is successful, that is, we manage to reach the ground state of $H_P$,  the probability
of finding the solution $\ket{\gamma_L}$ is $\frac{\eta^{2L+2}-\eta^{2L}}{\eta^{2L+2}-1}\sim 1-1/\eta^2$, 
which can be made very close to one with large $\eta$.  

The whole Hilbert space is of dimension $2^{n+L}$. However, our adiabatic operation (see next section)
will stay in the subspace of dimension $L+1$ spanned by $\ket{\gamma_\ell}$, where $H_B,H_P$
have the following matrix forms, 
\begin{equation}
H_B=
\begin{pmatrix}
0        &0         &\cdots      &\cdots         &0\\
0        &1         &0           &\cdots         &\vdots\\
\vdots   &0         &1           &\ddots         &\vdots\\
\vdots   &\ddots    &\ddots      &\ddots         &0\\
0        &\cdots    &\cdots      &0              &1
\end{pmatrix},
\label{mb}
\end{equation}
\begin{equation}
H_P=
\begin{pmatrix}
\frac{\eta}{2}&-\frac{1}{2}         &0                    &\cdots      &\cdots               &0\\
-\frac{1}{2}  &\frac{\eta+1/\eta}{2}                 &-\frac{1}{2}         &0           &\cdots               &\vdots\\
0             &-\frac{1}{2}         &\frac{\eta+1/\eta}{2}                 &-\frac{1}{2}&\ddots               &\vdots\\
\vdots        &\ddots               &\ddots               &\ddots      &\ddots               &\vdots\\
\vdots        &\cdots               &\ddots               &-\frac{1}{2}&\frac{\eta+1/\eta}{2}                 &-\frac{1}{2}\\
0             &\cdots               &\cdots               &0           &-\frac{1}{2}         &\frac{1}{2\eta}
\end{pmatrix}.
\label{mp}
\end{equation}

We can immediately construct a  quantum adiabatic algorithm with the following Hamiltonian
\begin{equation}
H(s)=(1-s)H_B+sH_P\,.
\label{ol}
\end{equation}
When $\eta=1$, this is the algorithm studied in Ref.~\cite{Dam}, which is polynomially slower than the corresponding quantum circuit algorithm. 
When $\eta >1$, this algorithm is exponentially slow as 
we can show rigorously that $H(s)$ has an exponentially small energy gap $\sim \eta^{-L}$
at $s^*=2/(\eta-1/\eta+2)$ (see Appendix A for details).  We will show in the next section how to avoid this small energy gap by introducing an intermediate Hamiltonian $H_I(t)$. 

\begin{figure}[!h]
	\centering
	\includegraphics[width=0.45\textwidth]{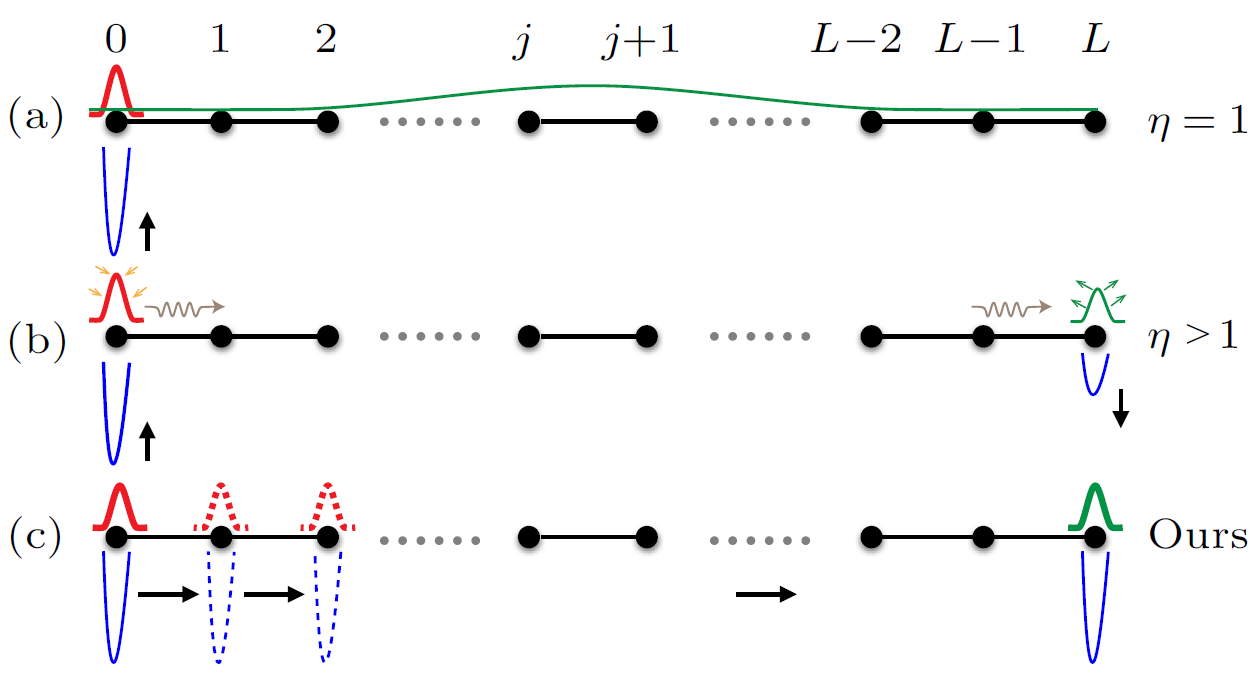}
	\caption{Schematic representations of three different quantum adiabatic algorithm.
	Each site of the lattice represents a quantum state $\ket{\gamma_j}$. Initially, a quantum particle resides in a potential well at site 0.
		(a) Algorithm  Eq.(\ref{ol}) with $\eta=1$: the potential well at site 0 slowly disappear and  the wave packet of 
		the particle spreads over the whole lattice. (b) Algorithm  Eq.(\ref{ol}) with $\eta>1$:  the potential well at site 0
		is lifted up slowly at site 0 while the other potential well is created  at site $L$ with
		increasing depth; during this process,  the particle tunnels from site 0 to site $L$. 
		(c) Our algorithm with an intermediate Hamiltonian: the potential well is moved adiabatically site by site while carrying the particle
		with it. }
	\label{fig:lattice}
\end{figure}

\section{Our adiabatic algorithm}
Before we present our adiabatic algorithm we first review the algorithm in Eq.(\ref{ol}) in an alternative perspective. 
As shown in Fig.\ref{fig:lattice}, we can construct a one dimensional lattice, where each site represents a quantum 
state $\ket{\gamma_j}$.   $H_B$ and $H_P$, as either diagonal or tridiagonal matrices (see Eqs.(\ref{mb},\ref{mp})),  
can be viewed as Hamiltonians defined on this lattice  $H_B$ represents a potential well at site 0. 
The diagonal elements of $H_P$ represent a potential that has two wells, one at site 0 and the other at site $L$ 
while its off-diagonal elements gives rise to hopping between lattice sites.  In this perspective, the adiabatically 
evolving Hamiltonian $H(s)$ in Eq.(\ref{ol}) is to move a particle initially residing in the potential well at site 0 to site $L$. 
When $\eta=1$, although $H_P$ has two  potential wells at sites 0 and  $L$, they are too shallow to hold bound states. 
As a result,  the end result of the adiabatic evolution is a wave packet spreading almost evenly over
the whole lattice. One has to repeat the process about $L$ times to find the particle at site $L$ by measurement~\cite{Dam}.  
This case is schematically shown in Fig.\ref{fig:lattice}(a). 

When $\eta>1$, the potential well at site 0 becomes shallower while the potential well at site $L$ gets deeper. 
The consequence is that $H_P$ has a bound state localized at site $L$ as described by Eq.(\ref{gs}). 
When $s$ changes slowly, the potential well of $H(s)$ at site 0  becomes shallower and the potential well at site $L$ 
becomes deeper. As the wells change their depths,   the particle initially at site 0 will tunnel to site $L$. 
As $H_P$ with $\eta >1$ has only one bound state,  one has
to change $s$ very slowly to keep the system in the ground state so that the particle will end up localized at site $L$. 
Physically, it is clear that this will 
become exponentially difficult as $L$ increases. This is captured mathematically by the exponentially small
gap at $s=s^*$. This case is shown in Fig.\ref{fig:lattice}(b).

Our algorithm is to generate a scenario depicted in Fig.\ref{fig:lattice}(c), where the potential well is moved slowly from site to site. 
We are able to find $\tau$, the time  spent moving the potential well from one site to the next, such that 
 $\tau$ is independent of the system size $L$ and at the same time $\tau$ is slow enough that the particle 
 moves with the potential well.  As a result, the time complexity of our algorithm is $O(L)$.  The scenario shown in 
 Fig.\ref{fig:lattice}(c) reminds us of the quantum tweezer proposed
in Ref.~\cite{Tweezer}.

If the lattice in Fig.\ref{fig:lattice} were replaced by a continuous line,  the scenario shown in Fig.\ref{fig:lattice}(c) could be  realized 
with the following  \sdg equation
\begin{equation}
\label{mse}
i \frac{\partial}{\partial t} \psi=-\frac{1}{2 m}\frac{\partial^2}{\partial x^2} \psi+V(x-vt)\psi
\end{equation}
where $V(x-vt)$ is a moving Gaussian potential well proportional to $-\exp[-(x-vt)^2]$. With  Galilean transformation, 
one can show that a particle initially in the ground state of $V(x)$ will remain in the ground state of $V(x-vt)$ at any time. 
That is, moving potential well $V(x-vt)$ will carry the particle with it. 
For a lattice, we only need to discretize the Hamiltonian in the above  \sdg equation.  We use $H_I(t)$ to denote 
the discretized Hamiltonian. Specifically,  the matrix of $H_I(t)$ are tridiagonal with
\begin{eqnarray}
&&(H_I)_{mm}(t)=\frac{1}{2\eta}+\frac{\eta}{2}[1-e^{-(t/\tau-m)^2}],\\
&&~~~~~~~~~~~~~~~~~~~~~~~~~~~~~~~~~~~~~(0\le m\le L) \nonumber\\
&&(H_I)_{m(m+1)}(t)=(H_I)_{(m+1)m}(t)=-\frac{1}{2}\,.\\
&&~~~~~~~~~~~~~~~~~~~~~~~~~~~~~~~~(0\le m\le L-1) \nonumber
\end{eqnarray}
where $\tau$ is a parameter independent of $L$ and $v=1/L\tau$. The detailed relation between $H_I(t)$ and the  \sdg equation (\ref{mse}) is given in Appendix B. 

Our algorithm is to use $H_I(t)$ as an intermediate Hamiltonian and construct three adiabatically changing Hamiltonians  
\begin{enumerate}[(i)]
\item  $H_1(s)=(1-s)H_B+sH_I(0)$ with $s$ changing slowly from 0 to 1; 
\item $H_I(t)$ for $0<t< L \tau$; 
\item $H_2(s)=(1-s)H_I(L\tau)+sH_P$ with $s$ changing slowly from 0 to 1.
\end{enumerate}
The algorithm works by preparing the system at state $\ket{\gamma_0}$ and  evolving it 
according to the above Hamiltonians one by one.

We can show that the minimum gaps of $H_1(s)$ and $H_2(s)$ are finite and independent of the system size $L$ (see Appendix A for detailed analysis).
This means that  the time spent with $H_1(s)$ and $H_2(s)$ is negligible when the system size $L$ is large enough. 
This allows us to focus on the evolution with $H_I(t)$. The minimum gap of $H_I(t)$ is also finite and independent of the system size $L$. After the evolution with $H_1(s)$, the system will evolve into 
a state very close to the ground state $\ket{\widetilde{\gamma}_1}$ of $H_I(0)$. 
Let us denote it as $\ket{\psi_1}=\ket{\widetilde{\gamma}_1}+\delta$,  where $|\delta|\ll 1$. If we evolve $\ket{\psi_1}$ with 
the continuous \sdg equation (\ref{mse}), with the Galilean transformation, we are sure that 
the system will stay very close to the ground state and the error $\delta$ will stay small. $H_I(t)$ 
is its discretized version. We show in Appendix B that during the evolution with $H_I(t)$ the error $\delta$ will also stay small.
As the evolution time with $H_I(t)$ is $L\tau$,  the time complexity of our algorithm is $O(L)$. 

It is interesting to compare our algorithm with algorithm Eq.(\ref{ol}) 
(or equivalently, scenario (b) and scenario (c) in Fig.\ref{fig:lattice}). 
Both algorithms have the same beginning Hamiltonian and the problem Hamiltonian, and employ the adiabatic process. 
However, their time complexities are profoundly different: our algorithm is exponentially faster. 
The crucial difference is due to the additional Hamiltonian $H_I(t)$. Alternatively, we can say that we have chosen a different adiabatic evolution path. This shows that one may exponentially speed up 
a quantum adiabatic algorithm by carefully designing an evolution path. 

There are infinitely many methods to construct intermediate Hamiltonians and, therefore, infinitely many ways to design an adiabatic
evolution path from the beginning Hamiltonian and the problem Hamiltonian. The simplest evolution path 
as in Eq.(\ref{ol}) is likely not efficient.  In our proof,  the intermediate Hamiltonian $H_I(t)$ 
was introduced to effectively turn on the terms ${\cal O}_{\ell}$ in $H_P$ one by one. 
This turns out to be exponentially efficient than Eq.(\ref{ol}), where  all terms in $H_P$ are  turned on simultaneously. 
However, as there are now more switching on and off, one has to do it very smoothly to suppress the error that may occur
during the switchings. It was pointed out in Ref.~\cite{zhang2014hierarchical} that any discontinuity in the derivatives 
of a switching function may lead to errors. 
The use of the Gaussian function in our proof (or algorithm) is to suppress this kind of error. 
In this perspective, our proof presents one possible effective way to
design the adiabatic evolution path.

\section{Conclusion}
\noindent
In this work we have presented a method to transform a quantum circuit algorithm to quantum adiabatic 
algorithm without loss of efficiency. This means that in principle designing an efficient quantum algorithm 
is now entirely a physical endeavor. Furthermore, our method gives  an analytical example to show 
that  some quantum adiabatic algorithm can have an exponential speedup with a properly chosen evolution path.

\section{acknowledgement}
This work was supported by the The National Key Research and Development Program of China (Grants No.~2017YFA0303302, No.~2018YFA030562) 
and the National Natural Science Foundation of China (Grants No.~11334001 and No.~11429402). \\

\appendix

\section{Analytical results of energy gaps}
In this Appendix we give detailed derivations of two mathematical results regarding minimum
energy gaps used in Section~\ref{sec:main}. We present these results in a self-contained manner
so that they can be read without knowing anything in our main text.

We define three $N\times N$  ($N\gg 1$) matrices, $\cb$, $\cp$, and $\cm$.
The matrix $\cb$ is diagonal with $\cb_{11}=0$ and $\cb_{mm}=1$ ($2\le m\le N$).
The matrix $\cp$ is tridiagonal with
\ba
&&\cp_{11}=\eta/2, \nonumber\\
&&\cp_{mm}=\frac{\eta+1/\eta}{2}\,,~~~~~(2\le m\le N-1) \nonumber\\
&&\cp_{NN}=\frac{1}{2\eta}, \\
&&\cp_{m(m+1)}=\cp_{(m+1)m}=-\frac{1}{2},~~~(1\le m\le N-1). \nonumber
\ea
And the matrix $\cm$ changes with time and is tridiagonal with
\ba
&&\cm_{mm}(t)=\frac{1}{2\eta}+\frac{\eta}{2}(1-e^{-(t/\tau-m+1)^2}),\nonumber\\
&&~~~~~~~~~~~~~~~~~~~~~~~~~~~~~~~(1\le m\le N) \nonumber\\
&&\cm_{m(m+1)}(t)=\cm_{(m+1)m}(t)=-\frac{1}{2},\\
&&~~~~~~~~~~~~~~~~~~~~~~~~~~~~(1\le m\le N-1) \nonumber
\ea
We let $\cm_0=\cm(0)$ and $\cm_f=\cm((N-1)\tau)$ for convenience.
\subsection{Exponentially small energy gap}
We consider  Hamiltonian  $\ch_a(s)=(1-s)\cb+s\cp$ with $s\in[0,1]$. We
shall show that for $\eta\ge 4$ the gap between the lowest two eigenvalues of this Hamiltonian
is  exponentially small as $N\rightarrow\infty$ at
\be
s^*=\frac{2}{\eta-1/\eta+2}\,.
\ee

At $s=s^*$, $\ch_a(s^*)$ can be written as
\begin{equation}
\ch_a(s^*)=s^*
\begin{pmatrix}
\frac{\eta}{2}&-\frac{1}{2}         &0                    &\cdots      &\cdots               &0\\
-\frac{1}{2}  &\eta                 &-\frac{1}{2}         &0           &\cdots               &\vdots\\
0             &-\frac{1}{2}         &\eta                 &-\frac{1}{2}&\ddots               &\vdots\\
\vdots        &\ddots               &\ddots               &\ddots      &\ddots               &\vdots\\
\vdots        &\cdots               &\ddots               &-\frac{1}{2}&\eta                 &-\frac{1}{2}\\
0             &\cdots               &\cdots               &0           &-\frac{1}{2}         &\frac{\eta}{2}
\end{pmatrix}.
\end{equation}
Since $s^*$ is a  constant independent of $N$, we can just discuss the gap
of $\ch^*_a=\ch_a(s^*)/s^*$.  Assume that $\ch^*_a$ has an eigenvalue $\lambda$ and  an eigenvector $X=(x_1,x_2,\cdots,x_N)^T$ 
that satisfy
\begin{equation}
\ch^*_a X=\lambda X\,.
\end{equation}
We write the above equation in its component form as
\begin{eqnarray}
\frac{\eta}{2}x_1-\frac{1}{2}x_2&=&\lambda x_1\nonumber\\
-\frac{1}{2}x_{k-1}+\eta x_k-\frac{1}{2}x_{k+1}&=&\lambda x_k,~~~~\\
-\frac{1}{2}x_{N-1}+\frac{\eta}{2} x_{N}&=&\lambda x_{N}\nonumber
\end{eqnarray}
where $k = 2,\cdots,N-1$. By  introducing two additional variable $x_0$ and $x_{N+1}$,
we can convert the above equations into the standard second order difference equation
\begin{equation}
x_{k-1}-2(\eta-\lambda)x_k+x_{k+1}=0\,,
\end{equation}
where $k=1,2,\cdots,N$ and  the boundary conditions are
\begin{equation}
x_0=\eta x_1,~~\eta x_{L}=x_{L+1}
\end{equation}
It has two types of solutions. Type I solution is given by
\begin{equation}
x_k=A \sin (k\alpha)+B \cos (k\alpha)\,,
\label{sin1}
\end{equation}
with $\lambda=\eta-\cos\alpha$. Type II solution is given by
\begin{equation}
x_k=A \sinh (k\alpha)+B \cosh (k\alpha)\,,
\label{sinh1}
\end{equation}
with $\lambda=\eta-\cosh\alpha$. The two boundary conditions determine the value of $\alpha$ and
$\lambda$. We are allowed to consider only the  situation $\alpha>0$.

For $\eta \ge 4$,  type I eigenvalue $\lambda=\eta-\cos\alpha>(\eta+1)/2$. However, according to the Gershgorin circle theorem, 
$\ch_a^*$ has and only has two eigenvalues smaller than $(\eta+1)/2$.
Therefore,  the smallest two eigenvalues are of type II.
For type II solution,  the boundary conditions are
\be
B=\eta(A \sinh \alpha +B \cosh \alpha) \,,
\ee
and
\ba
&&\eta(A \sinh N\alpha +B \cosh N\alpha)\nonumber\\
&=&A \sinh (N+1)\alpha +B \cosh (N+1)\alpha\,.
\ea
After eliminating $A$ and $B$ we have
\begin{equation}
\frac{\eta \sinh \alpha}{1-\eta\cosh \alpha}=\frac{\sinh(N+1)\alpha-\eta \sinh \alpha}{\eta \cosh N\alpha-\cosh (N+1)\alpha}\,,
\end{equation}
which can be simplified into
\begin{equation}
\eta^2 \sinh (N-1)\alpha-2\eta \sinh N\alpha+\sinh (N+1)\alpha=0\,.
\end{equation}
Let $z=e^\alpha$, we can rewrite the equation as follows
\begin{equation}
z^2-2\eta z +\eta^2-z^{-2(N-1)}(\eta-1/z)^2=0\,.
\end{equation}
As $\alpha>0$,  we have $z>1$. For convenience, we define
\begin{equation}
f(z)=z^2-2\eta z +\eta^2-\Delta^2\,,
\end{equation}
where $\Delta(z)=z^{-(N-1)}(\eta-1/z)$. Also note that in the following discussion we alway have $N\gg 1$ and $\eta\ge 4$.

It is easy to find $f(1)=0$, $f'(1)>0$ , $f(+\infty)>0$ , and $f(\eta)<0$. Moreover, $f(z)=0$ has at most 2 roots, for the $\ch_a^*$ has and only has 2 eigenvalues satisfying the equation~\eqref{sinh1}. Thus there is one root $z_1$ in the interval $(1,\eta)$ and $z_2$ in $(\eta,+\infty)$. 
At the same time, we can make $\Delta(z)$ arbitrarily small by increasing $N$. 
This implies that we can focus on the behavior of $f(z)$ near $z=\eta$. 

Consider another function $g(z)=z^2-2\eta z +\eta^2-\delta^2$, where $\delta$ is a positive constant. It is clear that $g(z)=0$ has two roots
$z_\pm=\eta\pm\delta$. As we can find an $N_0$ so that $\Delta<\delta$ for all $N>N_0$ and arbitrarily small $\delta$, 
the roots of $f(z)$, $z_1,z_2$,  are within the internal $(\eta-\delta,\eta+\delta)$ for $N>N_0$.
In other words, $|z_1-z_2|$ is no more than $2\delta$.  The energy gap $|\lambda_1-\lambda_2|$ is 
 bounded by the distance between $z_1$ and $z_2$ as
\begin{equation}
|\lambda_1-\lambda_2|=\frac{1}{2}|z_1-z_2+1/z_1-1/z_2|<\frac{1}{2}|z_1-z_2|\,.
\end{equation}
Thus we come into the conclusion: For $\eta\ge 4$, there exist a $N_0$ and for all $N>N_0$, the gap of $\ch_a^*$ is 
smaller than $O((\eta-\delta)^{-N})$, where $\delta>0$ is an arbitrarily small constant.

\subsection{The first finite gap}

Here we consider the energy gap for
\begin{equation}
\label{gap1}
\ch_b(s)=(1-s)\cb+s\cm_0\,.
\end{equation}
Here we present a simple proof that $\ch_b(s)$ has finite energy gap for the situation $\eta\ge5$.
It is convenient  to study the eigenvalue of
\begin{equation}
\tilde{\ch}_b(s)=\ch_b(s)/s=\frac{1-s}{s}(\cb-I)+\cm_0
\end{equation}
where $I$ is the identity matrix. \par

By using Gershgorin circle theorem, we can easily check that $\tilde{\ch_b}$ has an energy gap 
larger than $\eta(1-1/e)/2-3/2$, which is greater than 0 for $\eta\ge5$. 
Therefore, for $1/\eta<s\le1$, $\ch_b(s)$ has an energy gap larger than $(1-1/e)/2-3/2\eta$. For $0\le s\le1/\eta$, we can continue applying the Gershgorin circle theorem to $\ch_b(s)$ and find another gap lower bound $(3-1/e-5/\eta)/2$. Thus $\ch_b(s)$ has a $N$-independent gap between the smallest two eigenvalues.\par

\begin{figure}[!h]
	\centering
	\includegraphics[width=0.4\textwidth]{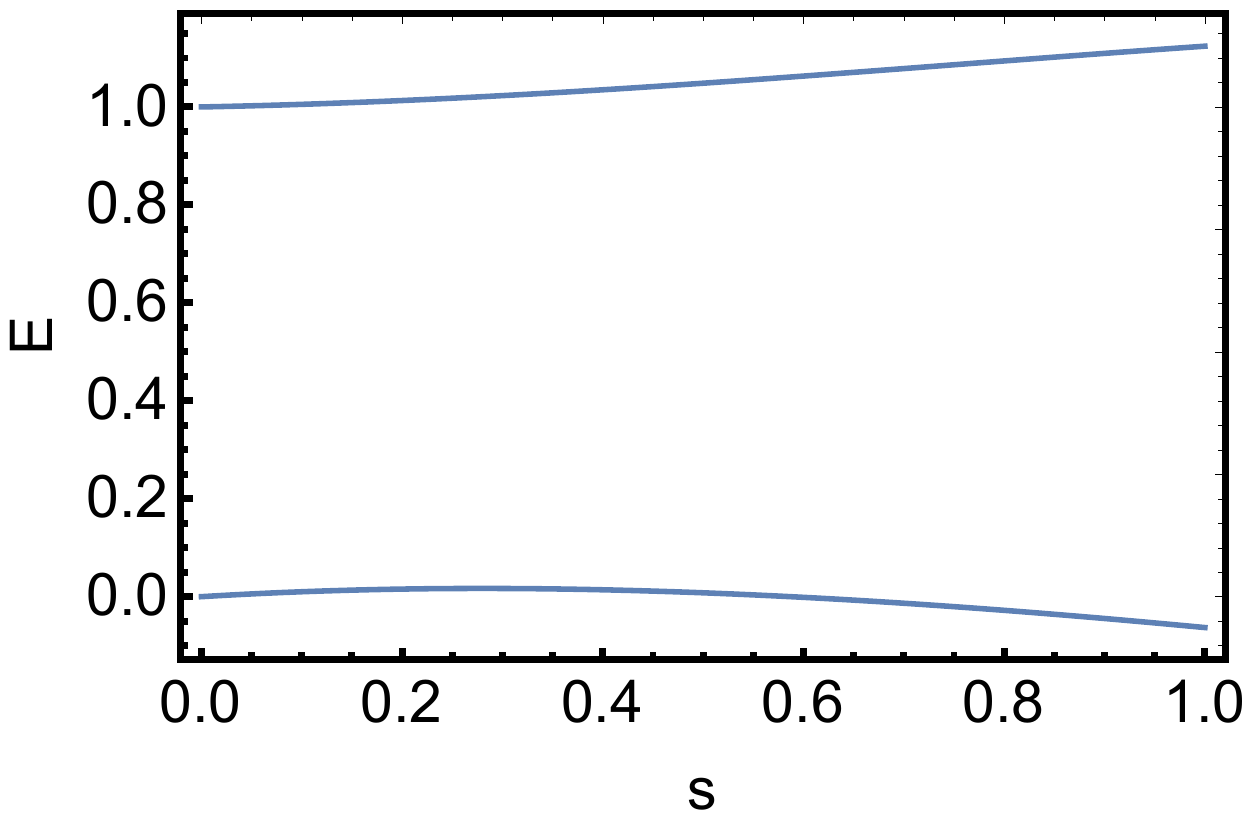}
	\caption{The lowest two energy levels of \eqref{gap1} at $\eta=4$ and $ N\rightarrow \infty$. }
	\label{fig:gap1}
\end{figure}

For $\eta<5$, the conclusion also holds true if $\eta$ is larger than a certain positive number, but the proof is rather complicated. Here we just show the numerical results in Fig.~\ref{fig:gap1} for the smallest two eigenvalues of \eqref{gap1} at $\eta=4$. In the discussion in the main text, 
$\eta$ is an arbitrary number larger than one. Therefore, if one has some doubts about the results here for  $\eta<5$, one can safely choose
$\eta\ge5$.

\subsection{The last finite gap}
We consider  the energy gap for
\begin{equation}
\label{gap2}
\ch_c(s)=(1-s)\cm_f+s\cp
\end{equation}
Note that \eqref{gap2} can be written as
\begin{equation}
\ch_c(s)-\cp=(s-1)[\cp-\cm_f]
\end{equation}
Let $\lambda_1$ and $\lambda_2$ be the two smallest eigenvalues of $\ch_c(s)$, and $\kappa_1$ and $\kappa_2$ 
the two smallest eigenvalues of $\cp$. From the main text,  we have already known that $\kappa_1=0$ and 
$\kappa_2>\frac{1}{2}(1/\eta+\eta)-1$.

It can be easily checked that the maximum and minimum of the eigenvalues of $\cp-\cm(LT)$ is $\eta/2e$ and $0$. With the Weyl's inequality, we have 
\begin{equation}
\lambda_1\le (1-s)\eta/2e<\eta/2e
\end{equation}
and 
\begin{equation}
\lambda_2\ge \kappa_2
\end{equation}
This implies that the upper bound of $\lambda_1$ is $\eta/2e$ and the lower bound of $\lambda_2$ is $\frac{1}{2}(1/\eta+\eta)-1$, which gives
\begin{equation}
\lambda_2-\lambda_1>\frac{1}{2}(1/\eta+\eta)-1-\frac{\eta}{2e}\,.
\end{equation}
This shows that the lowest energy gap of $\ch_c(s)$ is finite and independent of $N$ for $\eta\ge4$. Here we show the numerical results for the smallest two eigenvalues of \eqref{gap2} at $\eta=4$.
\begin{figure}[!h]
	\centering
	\includegraphics[width=0.4\textwidth]{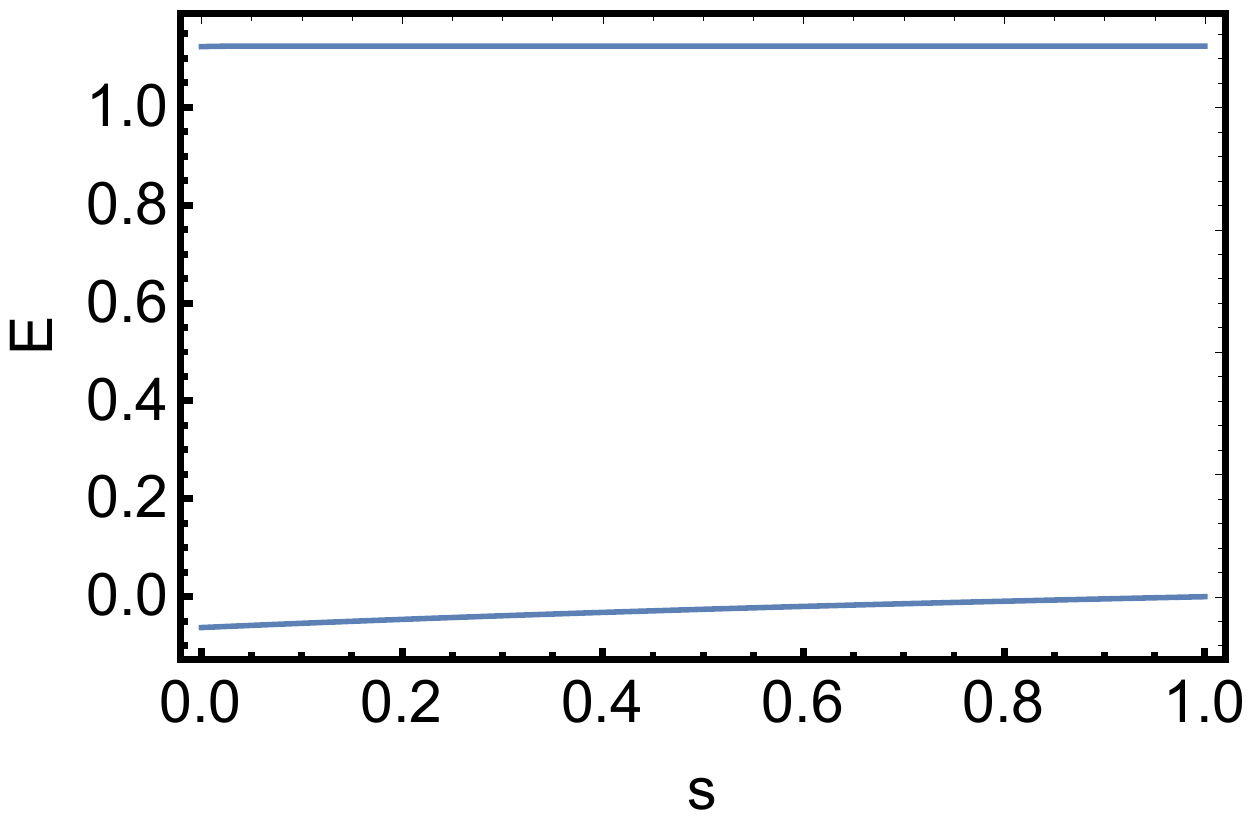}
	\caption{The lowest two energy levels of \eqref{gap2} for $\eta=4$ and $ N\rightarrow \infty$.}
	\label{fig:gap2}
\end{figure}
\section{Error analysis for $H_I(t)$}
In the main text, we considered two different but closely-related quantum dynamics. One is given by 
\begin{equation}
i \frac{d}{d t} U(t)=\cm(t) U(t)
\label{e}
\end{equation}
where $\cm(t)$ is the matrix form of $H_I(t)$ in the subspace spanned by $\ket{\gamma_\ell}$ and 
\begin{equation}
U(t)\equiv(u_0(t),u_1(t),\cdots,u_L(t))^T\,.
\end{equation}
Our goal is to prove that if $U(0)$ has a difference from the exact ground state $\tilde{\psi}(0)$ of $\cm(0)$, the difference will not grow too much when $t$ grows.\par
The other is given by 
\begin{equation}
i \frac{\partial}{\partial t} \psi(x,t)=-\frac{1}{2 m}\frac{\partial^2}{\partial x^2} \psi(x,t)+V(x-vt)\psi\,,
\label{ce}
\end{equation}
where $V(x-vt)$ is a moving Gaussian potential well and the wave function $\psi(x,t)$ is defined on 
the whole real axis. For this dynamical equation, by the argument of Galilean transformation, 
if $\psi(x,t)$ is initially the ground state for $V(x)$, $\psi(x,t)$ will stay in the ground state of 
$V(x-vt)$.  We  discretize Eq.\eqref{ce} as follows
\begin{equation}
\frac{\partial^2}{\partial x^2}\psi(x,t)\rightarrow \frac{\psi(x-h,t)+\psi(x+h,t)-2\psi(x,t)}{h^2}
\end{equation}
where $h=1/L$. When the circuit gate number $L$ is large, $h$ is a small interval of $x$. With such 
discretization, the two equations \eqref{ce} and \eqref{e} become identical to each other 
on the interval $x\in[0,1]$ when we set 
\begin{eqnarray}
m&=&1/h^2\nonumber\\
V(x-vt)&=&\frac{1}{2}(\eta+\frac{1}{\eta})-1-\frac{\eta}{2}e^{-(\frac{x}{h}-\frac{t}{\tau})^2},
\end{eqnarray}
where $v=h/\tau$. 

We are interested in how a small discrepancy between the initial states of Eq.\eqref{e} and Eq.\eqref{ce}
will grow. Specifically, let us define the discrepancy
\begin{equation}
e_m(t)=\psi(m h,t)-u_m(t)
\end{equation}
and let $E(t)=\{e_1(t),e_2(t),...\}^T$. We would like to know how large $E(t)$ can grow 
if initially $E(t)$ is small.  It is straightforward to check that  $E(t)$ satisfies
\begin{equation}
i \frac{\partial}{\partial t} E(t)=\cm(t) E(t)+\psi^{(4)}(t)\frac{ h^4}{24}\,,
\end{equation}
where 
\begin{equation}
\psi^{(4)}(t)\equiv\left(\frac{\partial ^4 \psi(h+\xi_1 h,t)}{\partial x^4},\frac{\partial ^4 \psi(2 h+\xi_2 h,t)}{\partial x^4},...\right)^T\,,
\end{equation}
where $\xi_m\in[0,1)$ are constants appearing in the remainders of Taylor expansions. 

\begin{figure}[!ht]
	\centering
	\includegraphics[width=0.4\textwidth]{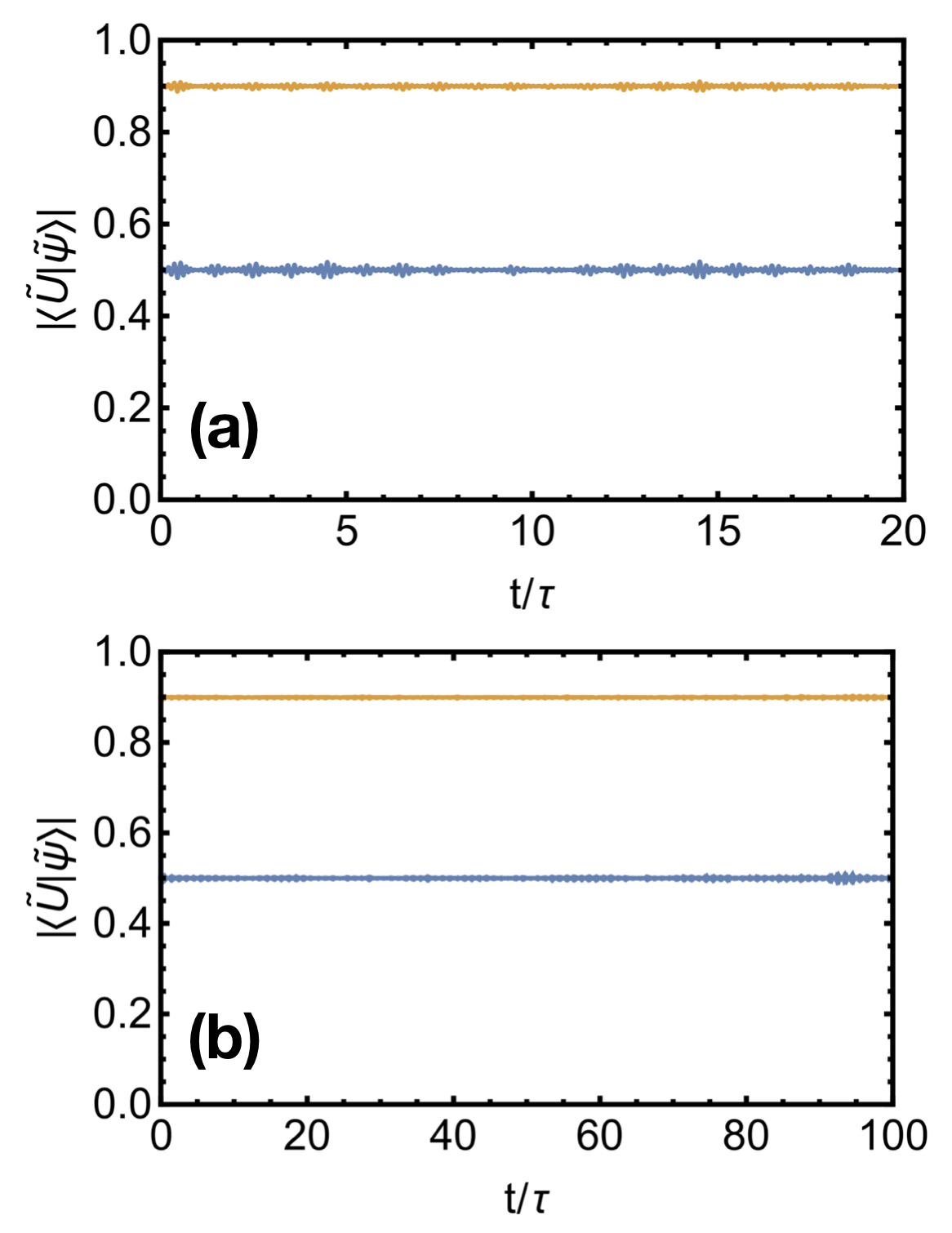}
	\caption{The error can also be described by $|\braket{\tilde{U}|\tilde{\psi}}|$, the product of the normalized actual state of \eqref{e} $\tilde{U}$ and $\cm(t)$'s exact ground state $\tilde{\psi}$. 
	The orange line and blue line indicate two initial conditions, where the initial similarity $|\braket{\tilde{U}|\tilde{\psi}}|=0.9$ and $0.5$. The similarity is almost unchanging with $t$ for both cases.
	In our computation, we set (a) $L=20$, $\tau=40$ and $\eta=4$; (b)  $L=100$, $\tau=40$ and $\eta=4$.}
	\label{fig:err}
\end{figure}

We focus on the situation that  $U(t)$ is initially close to the ground state $\psi(x,t)$. 
We can transform Eq. \eqref{ce} 
to a $h$-invariant form by rescaling $x=h s,v=h v_s$. 
\begin{equation}
i \frac{\partial}{\partial t} \phi(s,t)=-\frac{1}{2}\frac{\partial^2}{\partial s^2} \phi(s,t)+\tilde{V}(s-v_st)\phi(s,t)
\label{se}
\end{equation}
where $\phi(s,t)=\psi(x=hs,t)$.  The ground state can be written as 
\begin{equation}
\label{gt}
\phi(s,t)=\tilde{\phi}(s-v_s t) e^{i v_s s-i\frac{1}{2}v_s^2 t-i\varepsilon_0 t}\,,
\end{equation}
where $\varepsilon_0$ is the energy of the ground state. We replace $\psi^{(4)}$ with $\tilde{\phi}^{(4)}$
\begin{eqnarray}
\psi^{(4)}(t)&=& h^4e^{-i(\varepsilon_0+\frac{1}{2}v_s^2) t}\tilde{\phi}^{(4)}(t)\,,
\end{eqnarray}
where
\begin{equation}
\tilde{\phi}^{(4)}(t)\equiv\left(...,\frac{\partial ^4 \tilde{\phi}(j +\xi_j -v_st)}{\partial s^4}e^{i j v_s},...\right)^T\,.
\end{equation}
Then we get the ODE for the error $E(t)$
\begin{equation}\label{p}
i \frac{d}{d t} E(t)=\cm(t) E(t)+\frac{1}{24}\tilde{\phi}^{(4)}(t)e^{-i (\varepsilon_0 +\frac{1}{2}v_s^2)t}\,.
\end{equation}
The ODE can be directly solved as
\begin{eqnarray}
\label{int}
\tilde{E}(t)=\int_{0}^{L \tau} \frac{1}{24}e^{i\int^t[\cm(t')- (\varepsilon_0 +\frac{1}{2}v_s^2)]dt'}\phi^{(4)}(t)dt\,,
\end{eqnarray}
where
\begin{equation}
\tilde{E}(t)\equiv e^{i\int^t\cm(t')dt'}E(t)\,,
\end{equation}
$\tau=1/v_s$ is the evolution time for every step and $L$ is the gate number. We notice that
\begin{eqnarray}
|E(t)|^2&=&|\tilde{E}(t)|^2\nonumber\\
&=&|\int_{0}^{L\tau} \frac{1}{24}e^{i\int^t[\cm(t')- (\varepsilon_0 +\frac{1}{2}v_s^2)]dt'}\phi^{(4)}(t)dt|^2\nonumber\\
&\leq& \frac{1}{576}\int_{0}^{L\tau}|\phi^{(4)}(t)|^2dt
\end{eqnarray}
where $|.|^2$ is the 2-norm of a vector. With
\begin{eqnarray}
|\phi^{(4)}(t)|^2&=&\sum_m\left(\frac{\partial ^4 \tilde{\phi}(m+\xi_m-v_s t)}{\partial s^4}\right)^2\\
&\equiv&\sum_m g_m(v_s t)^2\,,
\end{eqnarray}
we have
\begin{eqnarray}
\int_{0}^{L\tau}|\phi^{(4)}(t')|^2 dt'&=&\sum_m \int_{0}^{L\tau}g_m(v_s t')^2 dt'\nonumber\\
&=&\sum_{m=0}^L \tau\int_{0}^{L}g_m(t)^2 dt\,.
\end{eqnarray}
It is evident that if $\int_{-\infty}^{\infty}|\partial ^4 \tilde{\phi}(s)/\partial s^4|^2ds$ converges then all $\int_{0}^{L}g_m(t)^2dt$ converge. This is true because 
\begin{equation}
\frac{\partial^2\tilde{\phi}}{\partial s^2}=2(\tilde{V}(s)-\varepsilon_0)\tilde{\phi}\,,
\end{equation}
where 
\begin{equation}
\tilde{V}(s)=\frac{1}{2}(\eta+\frac{1}{\eta})-1-\frac{\eta}{2}e^{-s^2}\,,
\end{equation}
and
\begin{equation}
\frac{\partial^4\tilde{\phi}}{\partial s^4}=\left[4(\tilde{V}(s)-\varepsilon_0)^2+2\frac{\partial^2\tilde{V}(s)}{\partial s^2}\right]\tilde{\phi}+4\frac{\partial\tilde{V}(s)}{\partial s}\frac{\partial\tilde{\phi}}{\partial s}\,.
\end{equation}
For a Gaussian potential $\tilde{V}(s)$, $4(\tilde{V}(s)-\varepsilon_0)^2+2\frac{\partial^2\tilde{V}(s)}{\partial s^2}$ has an upper bound, so the first term in the RHS is square-integrable. Besides, when $s$ is large, the ground state $\tilde{\phi}$ fades exponentially. With the Gaussian fades of $\partial \tilde{V}(s)/\partial s$, the second term in the RHS is also square-integrable. So $\partial ^4 \tilde{\phi}/\partial s^4$ is square-integrable. Therefore, all $\int_{0}^{L}g_m(t)^2dt$ will converge into a constant which is independent of the gate number $L$. So the error $|E(t)|^2$ is at most $O(L)$. Note that $U(t)$ is a discrete approximation of $\psi(x,t)$. As a result, it is normalized to
\begin{equation}
|U(t)|^2=\sum_{m=0}^L|u_m(t)|^2=L+1\,,
\end{equation}
the state of our quantum algorithm is $\tilde{U}=U/\sqrt{L+1}$. So what matters is the relative error $|E(t)|^2/|U(t)|^2$. This quantity will not grow with gate number $L$.\par

In the above discussion, we have discussed with the assumption that $U(t)$ is initially close to the ground state \eqref{gt} but we have not discussed how to achieve this. Eq.  \eqref{gt} is the ground state only when the dynamical \eqref{se} is defined on the entire real axis $s\in \mathbb{R}$ while $U(t)$ is defined only on the interval $s\in[0,L]$. In addition, the $\psi(x,t)$  slightly differs from exact ground state $\tilde{\psi}(t)$ of $\cm(t)$. We notice that the ground state of \eqref{se} is bounded near the potential well, and fades exponentially with $s$. So, when $s=t/\tau$ is far enough from 0 and $L$, Eq. \eqref{gt} will become a very good approximation and $\psi(x,t)$ is also very close to $\tilde{\psi}(x)$ after discretization and normalization. Therefore, when $s$ is far enough from 0 and $L$, the error between $\tilde{\psi}$ and $\tilde{U}$ will not grow with $t$.\par

Before $s=t/\tau$ evolves far enough from 0 or too close to $L$, we need a large $\tau$ to reduce the error between $\tilde{\psi}$ and $\tilde{U}$. The $\tau$ is independent of $L$ because $\cm(t)$ has a lower bound only related with $\eta$ for the ground energy gap. Since these two processes are only related with the first and last several qubits, once $\tau$ is set, 
the error will not change with the gate number $L$. Thus, the evolution from $H_I(0)$ to $H_I(L\tau)$ does bring error, but the error is controlled by $\tau$. This is confirmed by our  numerical result in Fig.\ref{fig:err} for different initial errors. In the figure, we have presented two sets of results, one for $L=20$ and the other for $L=100$, to show
that the error does not grow with $L$ (the size of the problem).

\end{document}